\documentclass[a4paper,11pt]{article}

\usepackage{contribution}
\usepackage{amssymb}
\usepackage{color,graphicx}
\usepackage{graphicx}
\usepackage{epsfig}

\newcommand{\weblink}[2][]{%
    \ifthenelse{\equal{#1}{}}%
    {\textnormal{\url{#2}}}%
    {\textnormal{\href{#2}{#1}}}%
}

\newcommand{\acknowledgements}[1]{%
  \bigskip\bigskip
  \textsf{\textbf{\Large Acknowledgements}} \\[2ex]
  {#1}
  \bigskip
}

\def\beq{\begin{equation}}
\def\eeq#1{\label{#1}\end{equation}}
\def\eeqn{\end{equation}}

\def\beqa{\begin{eqnarray}}
\def\eeqa#1{\label{#1}\end{eqnarray}}
\def\eeqan{\end{eqnarray}}

\let\bar=\overbar

\def\Dslash{\not{\hbox{\kern-4pt $D$}}}
\def\dslash{\not{\hbox{\kern-2pt $\del$}}}

\def\msb{{\bar{\ssstyle M \kern -1pt S}}}

\newcommand{\contribution}[7][]{%
  \clearpage
  \thispagestyle{plain}
  \ifthenelse{\equal{#1}{}}
  {\hypersetup{pdftitle={#2}}}
  {\hypersetup{pdftitle={#1}}}
  \hypersetup{pdfauthor={{#3} {#4}}}
  {\centering\normalfont\LARGE\bfseries\sffamily #2 \par\nobreak}
  \lhead{}
  \chead{%
    \textit{\footnotesize XIV International Conference on Hadron Spectroscopy
      (\weblink[\textit{hadron2011}]{http://www.hadron2011.de}), 13-17 June 2011, Munich, Germany}%
  }
  \rhead{}
  \bigskip
  \begin{center}
    {#3} {#4}\ifthenelse{\equal{#6}{}}{}{\footnote{\weblink[#6]{mailto:#6}}}
    \ifthenelse{\equal{#7}{}}{}{#7} \\
    \textit{#5}
  \end{center}
  \bigskip
}

\renewcommand{\abstract}[1]{%
  \begin{center}
    \begin{minipage}{0.85\textwidth}
      \begin{footnotesize}
        #1
      \end{footnotesize}
    \end{minipage}
  \end{center}
  \bigskip
}

\begin{document}

%
%
%
%
%
{  
\makeatletter
\@ifundefined{c@affiliation}%
{\newcounter{affiliation}}{}%
\makeatother
\newcommand{\affiliation}[2][]{\setcounter{affiliation}{#2}%
  \ensuremath{{^{\alph{affiliation}}}\text{#1}}}
\newcommand{\eq}{\begin{eqnarray}}
\newcommand{\en}{\end{eqnarray}}
\newcommand{\ra}{\rangle}
\newcommand{\la}{\langle}

%

\contribution[Mesons and baryons in the holographic soft-wall model]
{Mesons and baryons in the holographic soft-wall model}
{Valery E.}{Lyubovitskij}  
{\affiliation[  Institut f\"ur Theoretische Physik, 
                Universit\"at T\"ubingen, 
                Kepler Center for Astro,]{1} \\
                and Particle Physics,  Auf der Morgenstelle 14, 
                D-72076 T\"ubingen, Germany \\ 
 \affiliation[  Departamento de F\'\i sica y Centro Cient\'\i fico
                Tecnol\'ogico de Valpara\'\i so (CCTVal),]{2} \\
                Universidad T\'ecnica Federico Santa Mar\'\i a,
                Casilla 110-V, Valpara\'\i so, Chile}
{On leave of absence from Department of Physics, Tomsk State University,
634050 Tomsk, Russia} 
{\!\!$^,\affiliation{1}$, Thomas Gutsche\affiliation{1}, 
Ivan Schmidt\affiliation{2}, Alfredo Vega\affiliation{2}}

\abstract{
Mesons and baryons are considered in soft-wall holographic
approach based on the correspondence of string theory in AdS space
and conformal field theory in physical space-time.
The model generates Regge trajectories linear in $n$ and $J (L)$   
for the hadronic mass spectrum. Results obtained for heavy-light meson
masses and decay constants are consistent with predictions of HQET. 
In the baryon sector applications to the nucleon electromagnetic 
form factors and generalized parton distributions are discussed. 
}  

Based on the gauge/gravity duality~\cite{Maldacena:1997re}  
a class of AdS/QCD approaches was recently successfully developed 
for describing the phenomenology of hadronic properties. 
In order to break conformal invariance and
incorporate confinement in the infrared region two alternative
AdS/QCD backgrounds have been suggested in the literature: the
``hard-wall'' approach~\cite{Hard_wall1}, based on
the introduction of an IR brane cutoff in the fifth dimension, and
the ``soft-wall'' approach~\cite{Soft_wall1,Soft_wall2a,Soft_wall3a,%
SWplus,Soft_wall5,Soft_wall6,Soft_wall7,Soft_wall8,%
Soft_wall9,Gutsche:2011vb}, based on using a soft cutoff.  
In series of papers~\cite{Soft_wall5,Soft_wall8,Soft_wall9,Gutsche:2011vb}  
we developed the soft-wall approaches, which 
have been successfully applied for the study of meson and baryon properties.  
Here we present a summary of recent results: 
meson mass spectrum and decay constants of light and heavy mesons, 
nucleon electromagnetic form factors and generalized parton distributions.   
Our starting point are the 
effective $(d+1)$ dimensional actions formulated in AdS space 
in terms of boson or fermion bulk fields, which serve as holographic 
images of mesons and baryons. For illustration we consider the 
simplest actions --- for scalar fields ($J=0$)~\cite{Soft_wall8}
\eq 
S_0 = \frac{1}{2} \int d^dx dz \sqrt{g} e^{-\varphi(z)}
\biggl[ g^{MN} \partial_M S(x,z) \partial_N S(x,z)
- \Big(\mu_S^2 + \Delta V_0(z)\Big) \, S^2(x,z) \biggr] \,. 
\nonumber
\en
and $J=1/2$ fermions~\cite{Soft_wall7,Soft_wall9}: 
\eq
S_{1/2} &=&  \int d^dx dz \, \sqrt{g} \, e^{-\varphi(z)} \,
\biggl[ \frac{i}{2} \bar\Psi(x,z) \epsilon_a^M \Gamma^a
{\cal D}_M \Psi(x,z) - \frac{i}{2} ({\cal D}_M\Psi(x,z))^\dagger \Gamma^0
\epsilon_a^M \Gamma^a \Psi(x,z) \nonumber\\
&-& \bar\Psi(x,z) \Big(\mu_\Psi + \varphi(z)/R)\Big)\Psi(x,z)\biggr] \,, 
\nonumber 
\en 
where $S$ and $\Psi$ are the scalar and fermion bulk fields, 
${\cal D}_M$ is the covariant derivative acting on the fermion field, 
$\Gamma^a=(\gamma^\mu, - i\gamma^5)$ are the Dirac matrices, 
$\varphi(z) = \kappa^2 z^2$ is the dilaton field, $R$ is the AdS radius, 
$\Delta V_0(z)$ is the dilaton potential. 
$\mu_S$ and $\mu_\Psi$ are the masses of scalar and fermion bulk fields 
defined as $\mu_S^2 R^2 = \Delta_M (\Delta_M - d)$ and 
$\mu_\Psi R = \Delta_B - d/2$. Here 
$\Delta_M = \tau_M = 2 + L$ and $\Delta_B = \tau_B + 1/2 = 7/2 + L$
are the dimensions of scalar and fermion fields, which due to 
the QCD/gravity correspondence are related to the scaling dimensions 
(twists $\tau_M, \tau_B$) of the corresponding interpolating operators, 
where $L = {\rm max} \, | L_z |$~\cite{Soft_wall2a}.  
These actions give information about the propagation of bulk fields 
inside AdS space (bulk-to-bulk propagators), from inside 
to the boundary of the AdS space (bulk-to-boundary propagators) 
and bound state solutions - profiles of the Kaluza-Klein (KK) modes in 
extra-dimension, which correspond to the hadronic wave functions 
in impact space. We suppose a free propagation
of the bulk field along the $d$ Poincar\'e coordinates with four-momentum
$p$, and a constrained propagation along the $(d+1)$-th coordinate $z$
(due to confinement imposed by the dilaton field). 
In particular, it was shown~\cite{Soft_wall2a}   
that the extra-dimensional coordinate $z$ corresponds 
to the light-front impact variable. It was also shown~\cite{SWplus}  
that in case of the scattering problem the sign of the dilaton profile is 
important to fulfill certain model-independent constraints.  
But we recently showed~\cite{Gutsche:2011vb}, that in case of the 
bound state problem the sign of the dilaton profile is irrelevant, 
if the action is properly set up. Moreover, in solving the bound-state 
problem, it is more convenient to move the dilaton field from the exponential 
prefactor to the effective potential~\cite{Soft_wall2a,Gutsche:2011vb}.  
Then we use a KK expansion for the bulk fields factorizing 
the dependence on $d$ Poincar\'e coordinates $x$ and the holographic 
variable $z$. E.g. in case of scalar field it is given by    
$S(x,z) = \sum_n \ S_n(x) \ \Phi_{n}(z)$, 
where $n$ is the radial quantum number, $S_n(x)$
is the tower of the KK modes dual to
scalar mesons and $\Phi_n$ are their extra-dimensional profiles
(wave-functions) satisfying the Schr\"odinger-type equation with 
the potential depending on dilaton field. 
Then using the obtained wave functions $\Phi_n$ we calculate matrix elements 
describing hadronic processes. 
Finally, we present the results of our calculations for 
mesonic decay constants (Table 1) and spectrum (Tables 2 and 3), 
nucleon helicity-independent generalized parton distributions (GPDs) in Fig.1. 
Note, by construction we reproduce the power scaling of nucleon 
electromagnetic (EM) form factors at large $Q^2$ and our predictions for 
the EM radii are compare well with data: 
\eq 
\hspace*{-.2cm}
& &\la r^2_E \ra^p =  
0.91 \ {\rm fm}^2 \ {\rm (our)}\,,
\ 
0.77\ {\rm fm}^2 \ {\rm (data)}\, ; \ 
\la r^2_E \ra^n =    
- 0.12 \ {\rm fm}^2 \ {\rm (our)}\,, \ 
- 0.12 \ {\rm fm}^2 \ {\rm (data)}\,, \nonumber\\
\hspace*{-.2cm}
& &\la r^2_M \ra^p = 0.85 \ {\rm fm}^2 \ {\rm (our)}\,,
\ 
0.73 \ {\rm fm}^2 \ {\rm (data)}\, ; \  
\la r^2_M \ra^n =  0.88 \ {\rm fm}^2 \ {\rm (our)}\,,
\
0.76 \ {\rm fm}^2 \ {\rm (data)}\,. \nonumber
\en

\begin{center} 

Table 1. Decay constants $f_P$ (MeV) of pseudoscalar mesons. 

\def\arraystretch{.95}
\begin{tabular}{|l|c|c|}
\hline
Meson &Data & Our \\ \hline
$\pi^-$& $130.4\pm 0.03 \pm 0.2$ & 131 \\
\hline
$K^-$ & $156.1\pm 0.2 \pm 0.8$ & 155 \\
         \hline
$D^+$ & $206.7 \pm 8.9$ & 167 \\
\hline
$D_s^+$ & $257.5\pm6.1$ & 170 \\
\hline
$B^-$&$193\pm11$ & 139 \\
\hline
$B_s^0$&$253 \pm 8 \pm 7$ & 144 \\
\hline
\end{tabular}
\end{center}

\begin{center}

Table 2. Masses of light mesons

\def\arraystretch{.95}
\begin{tabular}{|l|c|c|c|l|l|l|l|}
\hline
Meson&$n$&$L$&$S$&\multicolumn{4}{c|}{Mass [MeV]} \\
\hline
$\pi$&0&0,1,2,3&0&$ 140$&$ 1355$ & $1777$&$2099$ \\ \hline
$\pi$&0,1,2,3&0&0&$140$&$1355$&$1777$&$2099$ \\ \hline
$K$& 0&0,1,2,3&$\;0\;$&$496$&$1505$&$1901$ & $2207$ \\ \hline
$f_0[\bar n n]$&0,1,2,3&1&1&$1114$&$1600$&$1952$&$2244$ \\ \hline
$f_0[\bar s  s]$&0,1,2,3&1&1&$1304$&$1762$&$2093$&$2372$ \\ \hline
$a_0(980)$&0,1,2,3&1&1&$1114$&$1600$&$1952$&$2372$ \\ \hline
$\rho(770)$&0,1,2,3&0&1&$804$&$1565$&$1942$&$2240$ \\ \hline
$\phi(1020)$ &0,1,2,3&0&1&$1019$&$1818$&$2170$&$2447$ \\ \hline
$a_1(1260)$&0,1,2,3&1&1&$1358$&$1779$&$2101$&$2375$ \\ \hline
\end{tabular}
\end{center}

\begin{center}

Table 3. Masses of heavy-light mesons

\def\arraystretch{.95}
\begin{tabular}{|l|c|c|c|c|c|c|c|c|}
\hline
Meson&$J^{\rm P}$&$n$&$L$&$S$&\multicolumn{4}{c|}{Mass [MeV]} \\
\hline
$D(1870)$&$0^{-}$&0&0,1,2,3         &0& 1857 & 2435 & 2696 & 2905 \\ \hline
$D^{\ast}(2010)$&$1^{-}$&0&0,1,2,3  &1& 2015 & 2547 & 2797 & 3000 \\ \hline
$D_s(1969)$&$0^{-}$&0&0,1,2,3       &0& 1963 & 2621 & 2883 & 3085 \\ \hline
$D^{\ast}_s(2107)$&$1^{-}$&0&0,1,2,3&1& 2113 & 2725 & 2977 & 3173 \\ \hline
$B(5279)$&$0^{-}$&0&0,1,2,3         &0& 5279 & 5791 & 5964 & 6089 \\ \hline
$B^{\ast}(5325)$&$1^{-}$&0&0,1,2,3  &1& 5336 & 5843 & 6015 & 6139 \\ \hline
$B_s(5366)$&$0^{-}$&0&0,1,2,3       &0& 5360 & 5941 & 6124 & 6250 \\ \hline
$B^{\ast}_s(5413)$&$1^{-}$&0&0,1,2,3&1& 5416 & 5992 & 6173 & 6298 \\ \hline
\end{tabular}
\end{center}

\vspace*{.2cm}
\begin{figure}[htb]
\begin{center}
\includegraphics[scale=0.55,angle=0]{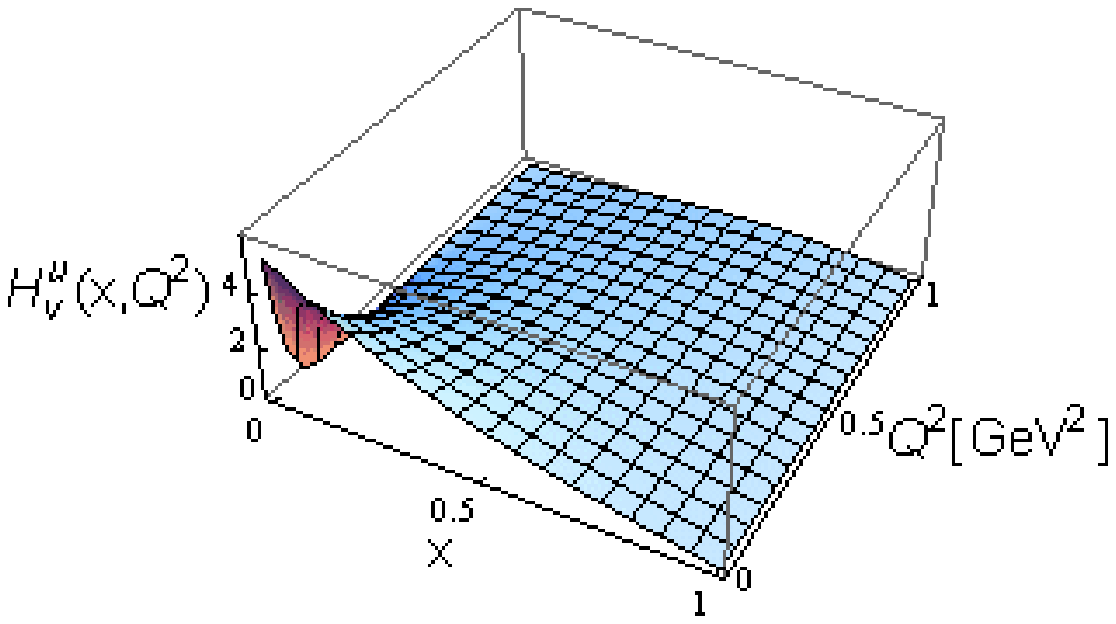} 
\includegraphics[scale=0.55,angle=0]{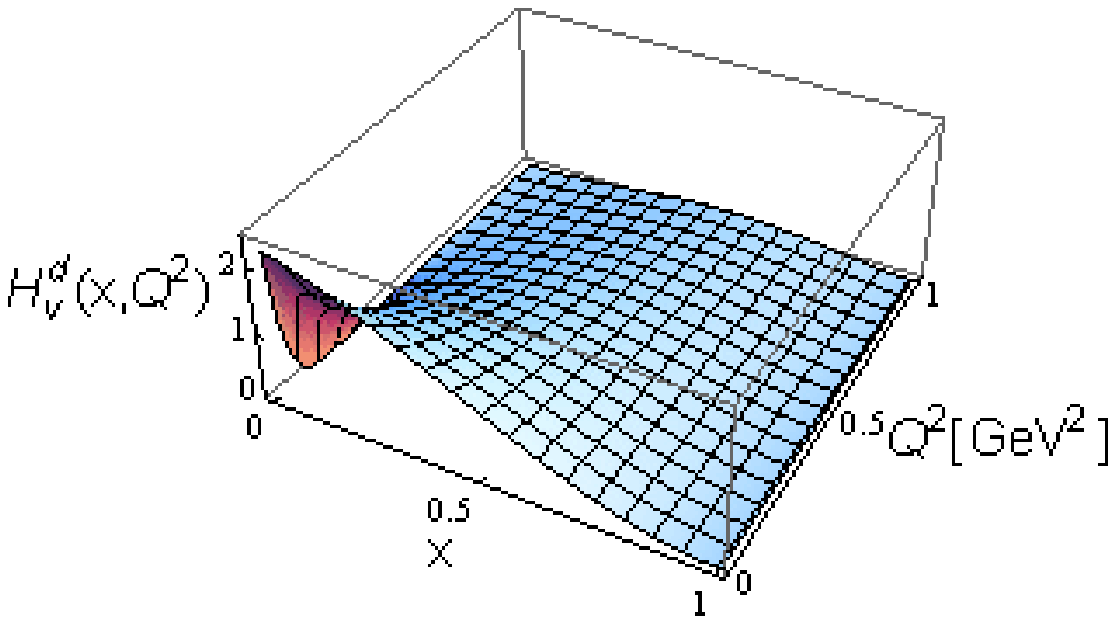} \\
\includegraphics[scale=0.55,angle=0]{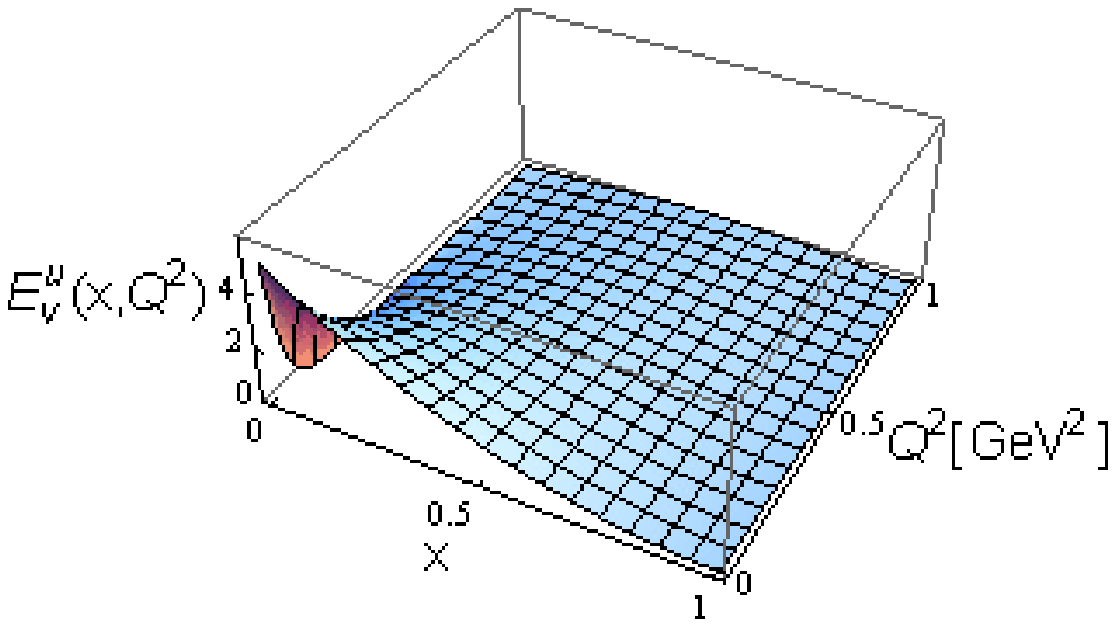} 
\includegraphics[scale=0.55,angle=0]{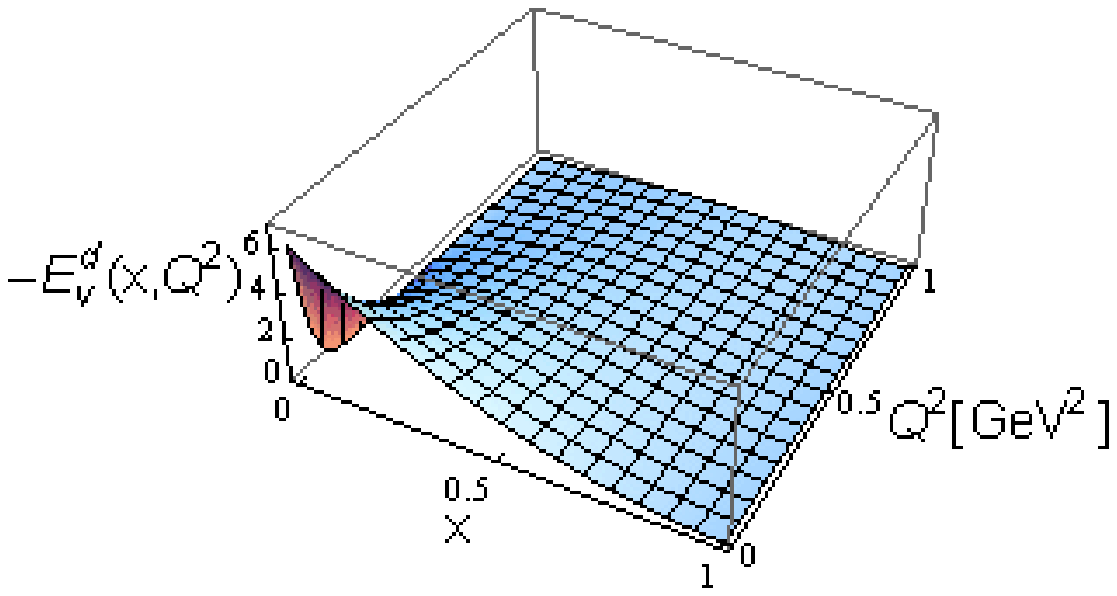} 
\caption{Nucleon helicity-independent GPDs in momentum space} 
\end{center}
\end{figure} 

\acknowledgements{
The authors thank Stan Brodsky and Guy de T\'eramond for useful discussions
and remarks. This work was supported by Federal Targeted Program ``Scientific
and scientific-pedagogical personnel of innovative Russia''
Contract No. 02.740.11.0238, by FONDECYT (Chile) under Grant No. 1100287. 
A.V. acknowledges the financial support from FONDECYT (Chile)
Grant No. 3100028. 
V.E.L. would like to thank Departamento de F\'\i sica y Centro
Cient\'\i fico Tecnol\'ogico de Valpara\'\i so (CCTVal), Universidad
T\'ecnica Federico Santa Mar\'\i a, Valpara\'\i so, Chile for warm
hospitality. 
}

}  


\end{document}